\documentclass[12pt]{article}
\usepackage{a4}
\usepackage{amsfonts}
\usepackage{amsmath}
\usepackage{latexsym}
\usepackage{amsfonts}
\usepackage{graphics}
\usepackage{graphicx}
\usepackage{epic}
\usepackage{eepic}

\begin{document}

\noindent hep-th/
   \hfill  March 2010 \\

\renewcommand{\theequation}{\arabic{section}.\arabic{equation}}
\thispagestyle{empty}
\vspace*{-1,5cm}
\noindent \vskip3.3cm

\begin{center}
{\Large\bf General trilinear interaction for arbitrary even higher spin gauge fields}

{\large Ruben Manvelyan ${}^{\dag\ddag}$, Karapet Mkrtchyan${}^{\dag\ddag}$ \\and Werner R\"uhl
${}^{\dag}$}
\medskip

${}^{\dag}${\small\it Department of Physics\\ Erwin Schr\"odinger Stra\ss e \\
Technical University of Kaiserslautern, Postfach 3049}\\
{\small\it 67653
Kaiserslautern, Germany}\\
\medskip
${}^{\ddag}${\small\it Yerevan Physics Institute\\ Alikhanian Br.
Str.
2, 0036 Yerevan, Armenia}\\
\medskip
{\small\tt manvel,ruehl@physik.uni-kl.de; karapet@yerphi.am}
\end{center}\vspace{2cm}

\bigskip
\begin{center}
{\sc Abstract}
\end{center}
\quad  Using Noether's procedure we present a complete solution for the trilinear interactions of arbitrary spins $s_{1},s_{2}, s_{3}$ in a flat background, and discuss the possibility to enlarge this construction to higher order interactions in the gauge field. Some classification theorems of the cubic (self)interaction with different numbers of derivatives and depending on relations between the spins are presented.  Finally the expansion of a general spin $s$ gauge transformation into powers of the field and the related closure of the gauge algebra in the general case are discussed.

\newpage

\section{Introduction and notations}
\quad The motivation for the investigation of higher spin gauge field (HSF) interactions can be summarized as the following list of the essential three points:
\begin{enumerate}
  \item The construction of interacting higher spin theories has been considered as an interesting task for itself and was always in the center of attention during the last thirty years \cite{MMR0}-\cite{review}.
  \item Particular attention arose during the last decade after discovering the holographic duality between the $O(N)$ sigma model in $d=3$ space and HSF gauge theory living in the space $AdS_4$ \cite{Kleb}. This case of holography is especially important  by the existence of two conformal points of the boundary theory and the possibility to describe them by the same HSF gauge theory with the help of \emph{spontaneous breaking of higher spin gauge symmetry and mass generation by a corresponding Higgs mechanism} \cite{Ruehl}-\cite{MR2}.
  \item Another still open point is verifying the holographic correspondence on the level of loop diagrams in the  general case, and the possibility to use this correspondence for real constructions of unknown local interacting theories on the bulk from more or less well known conformal field theories on the boundary.
\end{enumerate}
These complicated physical tasks necessitate quantum loop calculations for the HSF field theory \cite{MR3}, \cite{MR4},\cite{MR5} and therefore information about the manifest, off-shell and Lagrangian formulation of possible interactions for HSF. On the other hand one loop calculations are mainly interesting in the framework of their ultraviolet behaviour, when the difference between an $AdS$ and a flat space background can be neglected at least in the leading order. These motivations caused us during the last years to spend some effort on the construction of  possible couplings  which we started in a series of articles that involved couplings among different higher spin fields \cite{M, MMR, MR, MMR1} .
In our previous article \cite{MMR0} we directly construct a complete cubic selfinteraction for the case of spin $s=4$ in a flat background, and discuss the cubic selfinteraction for general spin $s$ with $s$ derivatives in the same background. The leading term of the latter interaction together with the gauge transformation of first order in the field was presented and investigated.

Here we turn to the trilinear interaction of  Fronsdal's \cite{Frons}  general spin $s_{1},s_{2},s_{3}$ gauge fields in a flat background (Section 2) and present the full solution of the corresponding Noether's equation (Section 3). Then we discuss a general classification theorem for (self)interactions based on our construction and the relation with other known couplings involving  (Weyl) tensors integrated into the interaction Lagrangians (Section 4). The last section 5 is devoted to a discussion of gauge transformations that are nonlinear in the gauge field, as they were invented in the classical paper \cite{vanDam2}, and the possibility to form Lie algebras of gauge transformations for even spin abelian gauge fields and in this context the construction of the fourth and higher order selfinteractions.

To handle these sections we should introduce here briefly our standard notations coming from our previous papers about HSF \cite{MMR0, M, MMR, MR, MMR1}. As usual we utilize instead of symmetric tensors such as $h^{(s)}_{\mu_1\mu_2...\mu_s}(z)$ the homogeneous polynomials in the vector $a^{\mu}$ of degree $s$ at the base point $z$
\begin{equation}
h^{(s)}(z;a) = \sum_{\mu_{i}}(\prod_{i=1}^{s} a^{\mu_{i}})h^{(s)}_{\mu_1\mu_2...\mu_s}(z) .
\end{equation}
Then we can write the symmetrized gradient, trace and divergence \footnote{To distinguish easily between "a" and "z" spaces we introduce the notation $\nabla_{\mu}$ for space-time derivatives $\frac{\partial}{\partial z^{\mu}}$.}
\begin{eqnarray}
&&Grad:h^{(s)}(z;a)\Rightarrow Gradh^{(s+1)}(z;a) = (a\nabla)h^{(s)}(z;a) , \\
&&Tr:h^{(s)}(z;a)\Rightarrow Trh^{(s-2)}(z;a) = \frac{1}{s(s-1)}\Box_{a}h^{(s)}(z;a) ,\\
&&Div:h^{(s)}(z;a)\Rightarrow Divh^{(s-1)}(z;a) = \frac{1}{s}(\nabla\partial_{a})h^{(s)}(z;a) .
\end{eqnarray}
Moreover we introduce the notation $*_a, *_b,\dots$ for a contraction in the symmetric spaces of indices $a$ or $b$
\begin{eqnarray}
  *_{a}&=&\frac{1}{(s!)^{2}} \prod^{s}_{i=1}\overleftarrow{\partial}^{\mu_{i}}_{a}\overrightarrow{\partial}_{\mu_{i}}^{a} .
   \label{0.12}
\end{eqnarray}
Then we see that the operators $(a\partial_{b}), a^{2}, b^{2}$ are dual (or adjoint) to $(b\partial_{a}),\Box_{a},\Box_{b}$ with respect to the "star" product of tensors with two sets of symmetrized indices  (\ref{0.12})
\begin{eqnarray}
    &&\frac{1}{n}(a\partial_{b})f^{(m-1,n)}(a,b)*_{a,b} g^{(m,n-1)}(a,b)= f^{(m-1,n)}(a,b)*_{a,b} \frac{1}{m}(b\partial_{a})g^{(m,n-1)}(a,b) ,\label{0.13}\quad\quad\\
   && a^{2}f^{(m-2,n)}(a,b)*_{a,b} g^{(m,n)}(a,b)=f^{(m-2,n)}(a,b)*_{a,b} \frac{1}{m(m-1)}\Box_{a} g^{(m,n)}(a,b) . \quad\quad\label{0.14}
\end{eqnarray}
In the same fashion gradients and divergences are dual with respect to the full scalar product in the space $(z,a,b)$
\begin{eqnarray}
  (a\nabla)f^{(m-1,n)}(z;a,b)*_{a,b} g^{(m,n)}(z;a,b) &=& -f^{(m-1,n)}(z;a,b)*_{a,b}\frac{1}{m}(\nabla\partial_{a}) g^{(m,n)}(z;a,b) .\nonumber\\ \label{1.15}
  \end{eqnarray}
Analogous equations can be formulated for the operators $b^{2}$ or $b\nabla$.

Here we will only present
Fronsdal's Lagrangian in terms of these conventions\footnote{From now on we will presuppose integration everywhere where it is necessary (we work with a Lagrangian as with an action) and therefore we will neglect all $d$ dimensional space-time total derivatives when making a partial integration.}:
\begin{equation}\label{1.42(1)}
 \mathcal{L}_{0}(h^{(s)}(a))=-\frac{1}{2}h^{(s)}(a)*_{a}\mathcal{F}^{(s)}(a)
    +\frac{1}{8s(s-1)}\Box_{a}h^{(s)}(a)*_{a}\Box_{a}\mathcal{F}^{(s)}(a) ,
\end{equation}
where $\mathcal{F}^{(s)}(z;a)$ is the Fronsdal tensor
\begin{eqnarray}
\mathcal{F}^{(s)}(z;a)=\Box h^{(s)}(z;a)-s(a\nabla)D^{(s-1)}(z;a) , \quad\label{0.32(1)}
\end{eqnarray}
and $D^{(s-1)}(z;a)$ is the deDonder tensor or traceless divergence of the higher spin gauge field
\begin{eqnarray}\label{0.42(D)}
 && D^{(s-1)}(z;a) = Divh^{(s-1)}(z;a)
-\frac{s-1}{2}(a\nabla)Trh^{(s-2)}(z;a) ,\\
&& \Box_{a} D^{(s-1)}(z;a)=0 .
\end{eqnarray}
The initial gauge variation of zeroth order in the spin $s$ field is
\begin{eqnarray}\label{0.5}
\delta_{(0)} h^{(s)}(z;a)=s (a\nabla)\epsilon^{(s-1)}(z;a) ,
\end{eqnarray}
with the traceless gauge parameter for the double traceless gauge field
\begin{eqnarray}
&&\Box_{a}\epsilon^{(s-1)}(z;a)=0 ,\label{0.6}\\
&&\Box_{a}^{2}h^{(s)}(z;a)=0 .
\end{eqnarray}
Therefore at this point we can see from (\ref{0.5}) and (\ref{0.6})  that the de Donder gauge condition
\begin{equation}\label{dd}
    D^{(s-1)}(z;a)=0 .
\end{equation}
is a correct generalization of the Lorentz gauge condition in the case of spin $s>2$.
Finally we note that in deDonder gauge (\ref{dd})  $\mathcal{F}^{(s)}(z;a)=\Box h^{(s)}(z;a)$  and the field $h^{(s)}$ decouples from it's trace in Fronsdal's Lagrangian (\ref{1.42(1)}).

\section{Noether's theorem in leading order}
\setcounter{equation}{0}
We consider three potentials $h^{(s_{1})}(z_{1};a), h^{(s_{2})}(z_{2};b), h^{(s_{3})}(z_{3};c)$ whose spins $s_{i}$ are assumed to be ordered
\begin{equation}
s_{1} \geq s_{2}\geq s_{3}
\end{equation}
For the interaction we make the cyclic ansatz
\begin{eqnarray}
&&\mathcal{L}_{I}^{(0,0)}(h^{(s_{1})}(a),h^{(s_{2})}(b),h^{(s_{3})}(c))=\sum_{n_{i}} C_{n_{1},n_{2},n_{3}}^{s_{1},s_{2},s_{3}} \int dz_{1}dz_{2}dz_{3} \delta (z_{3}-z_{1}) \delta(z_{2}-z_{1})\nonumber\\
&&\hat{T}(Q_{12},Q_{23},Q_{31}|n_{1},n_{2},n_{3})h^{(s_{1})}(z_{1};a)h^{(s_{2})}(z_{2};b)h^{(s_{3})}(z_{3};c)\label{1.2}
\end{eqnarray}
where
\begin{eqnarray}
\hat{T}(Q_{12},Q_{23},Q_{31}|n_{1},n_{2},n_{3})= (\partial_{a}\partial_{b})^{Q_{12}}(\partial_{b}\partial_{c})^{Q_{23}} (\partial_{c}\partial_{a})^{Q_{31}}(\partial_{a}\nabla_{2})^{n_{1}}(\partial_{b}\nabla_{3})^{n_{2}}( \partial_{c}\nabla_{1})^{n_{3}}\nonumber\\\label{1.3}
\end{eqnarray}
and the notation $(0,0)$ as a superscript means that it is an ansatz for terms without $Divh^{(s_{i}-1)}=\frac{1}{s_{i}}(\nabla_{i}\partial_{a_{i}})h^{(s_{i})}(a_{i})$ and $Trh^{(s_{i}-2)}=\frac{1}{s_{i}(s_{i}-1)}\Box_{a_{i}}h^{(s_{i})}(a_{i})$.
Denoting the number of derivatives by $\Delta$ we have
\begin{equation}
n_{1}+n_{2}+n_{3} = \Delta \label{delta}
\end{equation}
We shall later determine the range of possible values of $\Delta$. As balance equations we have
\begin{eqnarray}
n_{1}+Q_{12}+Q_{31} = s_{1} \nonumber\\
n_{2}+Q_{23}+Q_{12} = s_{2} \nonumber\\
n_{3} + Q_{31} + Q_{23} = s_{3}\label{1.6}
\end{eqnarray}
These equations are solved by
\begin{eqnarray}\label{q}
Q_{12} = n_{3}-\nu_{3}\nonumber\\
Q_{23} = n_{1} - \nu_{1} \nonumber\\
Q_{31} = n_{2} - \nu_{2}
\end{eqnarray}
Since the l.h.s. cannot be negative, we have
\begin{equation}
n_{i} \geq  \nu_{i}
\end{equation}
The $\nu_{i}$ are determined to be
\begin{equation}
\nu_{i} = 1/2 (\Delta +s_{i} -s_{j} -s_{k}), \quad i,j,k \quad  \textnormal{are all different}
\end{equation}
These $\nu_{i}$ must also be nonnegative, since otherwise the natural limitation of the $Q_{ij}$ to nonnegative values
would imply a boundary value problem which has only a trivial solution (see below).
It follows that the minimal possible $\Delta$ is expressed by Metsaev's  (see \cite{Metsaev} equ. (5.11)-(5.13)) formula (using the ordering of the $s_{i}$).
\begin{equation}
\Delta_{min} = \max{[s_{i} +s_{j} -s_{k}]} = s_{1}+ s_{2} -s_{3}\label{2.25}
\end{equation}
For example
\begin{equation}
\Delta_{min} = 6 \quad \textnormal{for}\quad s_{1}=s_{2} = 4, s_{3} =2
\end{equation}
This value and the ordering of the $s_{i}$ implies for the $\nu_{i}$
\begin{eqnarray}
\nu_{1}= s_{1}-s_{3} \nonumber\\
\nu_{2} =s_{2}-s_{3} \nonumber\\
\nu_{3} = 0
\end{eqnarray}

We use Noether's theorem to derive recursion relations which are then solved. By variation w.r.t.
$h^{(s_{i})}$ we obtain three currents whose divergences must vanish on shell. We need only do the explicit variation once:
\begin{eqnarray}
J^{(3)}(z_{3};c) = \sum C_{n_{1},n_{2},n_{3}}^{s_{1},s_{2},s_{3}}\int dz_{1}dz_{2}\delta(z_{3}-z_{1})\delta(z_{3}-z_{2}) \nonumber \\
(\partial_{a}\partial_{b})^{Q_{12}} (\partial_{b} c)^{Q_{23}}
(c \partial)^{Q_{31}} (\partial_{a}\nabla_{2})^{n_{1}}(\partial_{b}\nabla_{3})^{n_{2}}(c\nabla_{1})^{n_{3}} \nonumber\\
h^{(s_{1})}(z_{1};a) h^{(s_{2})}(z_{2};b)
\end{eqnarray}
having the divergence
\begin{eqnarray}
(\partial_{c}\nabla_{3})J^{(3)}(z_{3};c) = \sum C_{n_{1},n_{2},n_{3}}^{s_{1},s_{2},s_{3}} \nonumber\\
\{n_{3} (\nabla_{1}\nabla_{3}) (\partial_{a}\partial_{b})^{Q_{12}}(\partial_{b}c)^{Q_{23}} (c\partial_{a})^{Q_{31}}(\partial_{a}\nabla_{2})^{n_{1}}
(\partial_{b}\nabla_{3})^{n_{2}}(c\nabla_{1})^{n_{3}-1} \nonumber\\
+Q_{23}(\partial_{a}\partial_{b})^{Q_{12}}(\partial_{b}c)^{Q_{23}-1} (c\partial_{a})^{Q_{31}}
(\partial_{a}\nabla_{2})^{n_{1}}(\partial_{b}\nabla_{3})^{n_{2}+1}(c\nabla_{1})^{n_{3}}\nonumber\\
+Q_{31}(\partial_{a}\partial_{b})^{Q_{12}}(\partial_{b}c)^{Q_{23}}(c\partial_{a})^{Q_{31}-1} (\partial_{a}\nabla_{2})^{n_{1}}(\partial_{a}\nabla_{3})
(\partial_{b}\nabla_{3})^{n_{2}}(c\nabla_{1})^{n_{3}}\}\nonumber\\
h^{(s_{1})}(z_{1};a)h^{(s_{2})}(z_{2};b) \mid z_{1}=z_{2}=z_{3} \label{ddd}
\end{eqnarray}
This divergence (and the corresponding divergences of the currents $J^{(1,2)}$) must vanish on shell.

We shall develop now a recursive algorithm. First we study the terms not containing any deDonder expression
$D^{(s_{i}-1)}, i=1,2,3$:
\begin{equation}
D^{(s_i -1)} = \frac{1}{s_i}[(\partial_{a_{i}}\nabla_i) -1/2 (a_{i}\nabla_i)\Box_{a_{i}}] h^{(s_i)}(z_{i};a_{i}) , \quad a_{i}=a,b,c.
\end{equation}
We use that
\begin{equation}
(\nabla_{1}\nabla_{3}) = 1/2 [\Box_{2} - \Box_{1} -\Box_{3} ]\label{1.17}
\end{equation}
and
\begin{eqnarray}
  \Box_i h^{(s_i)}(z_{i};a_{i}) &=& \mathcal{F}^{(s_{i})}(z_{i};a_{i})+ s_i (a_{i}\nabla)D^{(s_i -1)}\label{1.18}\\
   \Box_i \epsilon^{(s_i-1)}(z_{i};a_{i})  &=& \delta^{(0)}_{i}D^{(s_i -1)}\label{1.19}
\end{eqnarray}
where $ \mathcal{F}^{(s_{i})}(z_{i};a_{i})$ is Fronsdal's gauge invariant equation of motion and can be dropped on shell. So the $n_{3}$-term of (\ref{ddd}) does not contribute to the leading order terms.
On the other hand the $Q_{23}$-term is purely leading order. The $Q_{31}$-term contains
\begin{equation}
(\partial_{a}\nabla_{3}) = - (\partial_{a}\nabla_{2}) -(\partial_{a}\nabla_{1})\label{1.20}
\end{equation}
Only the first term yields a leading order contribution, the next one is a divergence term. A possibility to classify the higher order terms is to count the divergence and the deDonder operators separately, say by numbers $m_{1},m_{2}$ respectively.

In the leading order (l. o.) terms we renumber the powers $n_{1}\rightarrow n_{1}+1$ in the $Q_{23}$-term and $n_{2}\rightarrow n_{2}+1$ in the l.o. $Q_{31}$ term. We get
\begin{eqnarray}
[(n_{1}+1-\nu_{1})C_{n_{1}+1,n_{2},n_{3}}^{s_{1},s_{2},s_{3}}- (n_{2}+1-\nu_{2})C_{n_{1},n_{2}+1,n_{3}}^{s_{1},s_{2},s_{3}}]\label{1.21}\\
(\partial_{a}\partial_{b})^{n_{3}-\nu_{3}}(\partial_{b}c)^{n_{1}-\nu_{1}}(c\partial_{a})^{n_{2}-\nu_{2}}
(\partial_{a}\nabla_{2})^{n_{1}+1}(\partial_{b}\nabla_{3})^{n_{2}+1}(c\nabla_{1})^{n_{3}} = 0\nonumber
\end{eqnarray}
It follows that the factor in the square bracket must vanish. Two analogous relations follow from the two other currents.
The solution of these three recursion relations is
\begin{equation}
C_{n_{1},n_{2},n_{3}}^{s_{1},s_{2},s_{3}} = const \quad {\sum n_{i}-\sum \nu_{i}\choose  n_{1}-\nu_{1},n_{2}-\nu_{2},n_{3}-\nu_{3}} \label{1.22}
\end{equation}
Comparison with (\ref{q}) proves that we can present the trinomial coefficient also as
\begin{equation}
C_{Q_{12},Q_{23},Q_{31}}^{s_{1},s_{2},s_{3}} = const\quad {\sum n_{i}-\sum \nu_{i} \choose Q_{12},Q_{23},Q_{31}}\label{1.23}
\end{equation}
We see that the number of contractions between indices of our three fields $Q_{12},Q_{23},Q_{31}$ define our interaction completely.

Finally we want to make a remark concerning the case where two or all three of these fields are equal. Then we get only two or one current whose divergences vanish on shell. But in this case we have a symmetry which restores the result (\ref{1.21}), (\ref{1.22}) and shows that this is correct in all cases.

\section{Cubic interactions for arbitrary spins: Complete solution of the Noether's procedure}
\setcounter{equation}{0}

To derive the next terms of interaction containing one deDonder expression we turn to the Lagrangian formulation of the task and solve Noether's equation
\begin{equation}\label{2.24}
\sum^{3}_{i=1}\delta^{(1)}_{i}\mathcal{L}^{0}_{i}(h^{(s_{i})}(a))
+\sum^{3}_{i=1}\delta^{(0)}_{i}\mathcal{L}_{I}(h^{(s_{1})}(a),h^{(s_{2})}(b),h^{(s_{3})}(c))=0 ,
\end{equation}
where
\begin{eqnarray}
  \delta^{(0)}_{i}h^{(s_{i})}(a_{i}) &=& s_{i}(a_{i}\nabla_{i})\epsilon^{s_{i}-1}(z_{i};a_{i}) \\
  \mathcal{L}^{0}_{i}(h^{(s_{i})}(a))&=& -\frac{1}{2}h^{(s_{i})}(a_{i})*_{a_{i}}\mathcal{F}^{(s_{i})}(a_{i})
    +\frac{1}{8s_{i}(s_{i}-1)}\Box_{a_{i}}h^{(s_{i})}(a_{i})*_{a_{i}}\Box_{a_{i}}\mathcal{F}^{(s)}(a_{i})\nonumber\\
\end{eqnarray}
Shifting $\delta^{(1)}_{i}$ by a trace term in the same way as in \cite{MMR0} we obtain the following functional equation:
\begin{eqnarray}
 &&\sum^{3}_{i=1}\delta^{(0)}_{i}\mathcal{L}_{I}(h^{(s_{1})}(a),h^{(s_{2})}(b),h^{(s_{3})}(c))=0 + O(\mathcal{F}^{(s_{i})}(a_{i}))\label{2.28}
\end{eqnarray}
We solve this equation starting from the ansatz (\ref{1.2}), (\ref{1.3}) and integrating level by level in means of its dependence on deDonder tensors and traces of higher spin gauge fields.

Actually we have to solve the following equation:
\begin{eqnarray}
   &&C^{\{s_i\}}_{\{n_i\}}\hat{T}(Q_{ij}|n_{i}) [(a\nabla_{1})\epsilon^{(s_{1}-1)}h^{(s_{2})}h^{(s_{3})}
   +h^{(s_{1})}(b\nabla_{2})\epsilon^{(s_{2}-1)}h^{(s_{3})}+h^{(s_{1})}h^{(s_{2})}(c\nabla_{3})\epsilon^{(s_{3}-1)}]\nonumber\\
   &&=0 +O(\mathcal{F}^{(s_{i})}(a_{i}))
\end{eqnarray}
Taking into account that due to (\ref{1.6})
\begin{equation}\label{2.30}
  \hat{T}(Q_{ij}|n_{i})(a_{i}\nabla_{i})\epsilon^{(s_{i}-1)}(a_{i})=[\hat{T}(Q_{ij}|n_{i}),(a_{i}\nabla_{i})]\epsilon^{(s_{i}-1)}(a_{i})
\end{equation}
we see that all necessary information for the recursion can be found calculating these commutators
\begin{eqnarray}
  [\hat{T}(Q_{ij}|n_{i}),(a\nabla_{1})]&=&Q_{31}\hat{T}(Q_{12},Q_{23},Q_{31}-1|n_{1}, n_{2}, n_{3}+1)\nonumber\\
  &&-Q_{12}\hat{T}(Q_{12}-1,Q_{23},Q_{31}|n_{1}, n_{2}+1, n_{3})\nonumber\\
  &&+n_{1}\hat{T}(Q_{12},Q_{23},Q_{31}|n_{1}-1, n_{2}, n_{3})(\nabla_{1}\nabla_{2})\quad\quad\nonumber\\
  &&-Q_{12}\hat{T}(Q_{12}-1,Q_{23},Q_{31}|n_{1}, n_{2}, n_{3})(\partial_{b}\nabla_{2}),\label{2.31}
\end{eqnarray}
\begin{eqnarray}
  [\hat{T}(Q_{ij}|n_{i}),(b\nabla_{2})]&=&Q_{12}\hat{T}(Q_{12}-1,Q_{23},Q_{31}|n_{1}+1, n_{2}, n_{3})\nonumber\\
  &&-Q_{23}\hat{T}(Q_{12},Q_{23}-1,Q_{31}|n_{1}, n_{2}, n_{3}+1)\nonumber\\
  &&+n_{2}\hat{T}(Q_{12},Q_{23},Q_{31}|n_{1}, n_{2}-1, n_{3})(\nabla_{2}\nabla_{3})\quad\quad\nonumber\\
  &&-Q_{23}\hat{T}(Q_{12},Q_{23}-1,Q_{31}|n_{1}, n_{2}, n_{3})(\partial_{c}\nabla_{3}),\label{2.32}
\end{eqnarray}
\begin{eqnarray}
  [\hat{T}(Q_{ij}|n_{i}),(c\nabla_{3})]&=&Q_{23}\hat{T}(Q_{12},Q_{23}-1,Q_{31}|n_{1}, n_{2}+1, n_{3})\nonumber\\
  &&-Q_{31}\hat{T}(Q_{12},Q_{23},Q_{31}-1|n_{1}+1, n_{2}, n_{3})\nonumber\\
  &&+n_{3}\hat{T}(Q_{12},Q_{23},Q_{31}|n_{1}, n_{2}, n_{3}-1)(\nabla_{3}\nabla_{1})\quad\quad\nonumber\\
  &&-Q_{31}\hat{T}(Q_{12},Q_{23},Q_{31}-1|n_{1}, n_{2}, n_{3})(\partial_{a}\nabla_{1}),\label{2.321}
\end{eqnarray}
where we used relations like (\ref{1.18}) and (\ref{1.20}).
In these commutators we can use also the following identities
\begin{eqnarray}
  \nabla_{1}\nabla_{2}&=&\frac{1}{2}(\Box_{3}-\Box_{2}-\Box_{1}),\nonumber\\
  \nabla_{2}\nabla_{3}&=&\frac{1}{2}(\Box_{1}-\Box_{2}-\Box_{3}),\nonumber\\
  \nabla_{3}\nabla_{1}&=&\frac{1}{2}(\Box_{2}-\Box_{3}-\Box_{1}).
\end{eqnarray}
Now we see immediately from the first two lines of (\ref{2.31})-(\ref{2.321}) that these contribute to (\ref{2.28}) as leading order terms and yield the same equations for the $C^{s_{i}}_{n_i}$ coefficients as (\ref{1.21})
\begin{eqnarray}
  && (Q_{31}+1)C_{n_{1},n_{2}+1,n_{3}}^{s_{1},s_{2},s_{3}}- (Q_{12}+1)C_{n_{1},n_{2},n_{3}+1}^{s_{1},s_{2},s_{3}}=0 \label{2.33}\\
  && (Q_{12}+1)C_{n_{1},n_{2},n_{3}+1}^{s_{1},s_{2},s_{3}}- (Q_{23}+1)C_{n_{1}+1,n_{2},n_{3}}^{s_{1},s_{2},s_{3}}=0 \label{2.34}\\
  && (Q_{23}+1)C_{n_{1}+1,n_{2},n_{3}}^{s_{1},s_{2},s_{3}}- (Q_{31}+1)C_{n_{1},n_{2}+1,n_{3}}^{s_{1},s_{2},s_{3}}=0\label{2.35}
\end{eqnarray}
with the solution (\ref{1.22}) or (\ref{1.23}).

To find the full interaction we follow the same strategy as in the case $s=4$  \cite{MMR0} and introduce the following classification for the higher order interaction terms in $D$ and $\bar{h}=Trh$ :
\begin{equation}\label{intlag}
    \mathcal{L}_{I}=\sum_{i,j=0,1,2,3 \atop  i+j\leq 3} \mathcal{L}_{I}^{(i,j)}(h^{(s)}) ,
\end{equation}
where
\begin{equation}\label{ij}
    \mathcal{L}_{I}^{(i,j)}(h^{(s)}) \sim \nabla^{s-i} (D)^{i} (\bar{h}^{(s)})^{j} (h^{(s)})^{3-j-i} .
\end{equation}
In this notation the leading term described in the second section is $\mathcal{L}_{I}^{(0,0)}(h^{(s)})$.

To integrate Noether's equation next to the leading term we have to insert in (\ref{2.28}) the last two lines of (\ref{2.31})-(\ref{2.321}) and use two important relations (\ref{1.18}), (\ref{1.19}). Thus we arrive at the following $O(D)$ solution:
\begin{eqnarray}
  \mathcal{L}_{I}^{(1,0)}&=&\sum_{n_{i}} C_{n_1,n_2,n_3}^{s_1,s_2,s_3} \int dzdz_{1}dz_{2}dz_{3} \delta(z_{1}-z)\delta (z_{3}-z)\delta(z_{2}-z)\nonumber\\
  \Big[&+&\frac{s_{1}n_{1}}{2}\hat{T}(Q_{ij}|n_{1}-1,n_{2},n_{3})D^{(s_{1}-1)}h^{(s_{2})}h^{(s_{3})}\nonumber\\
  &+&\frac{s_{2}n_{2}}{2}\hat{T}(Q_{ij}|n_{1},n_{2}-1,n_{3})h^{(s_{1})}D^{(s_{2}-1)}h^{(s_{3})}\nonumber\\
  &+&\frac{s_{3}n_{3}}{2}\hat{T}(Q_{ij}|n_{1},n_{2},n_{3}-1)h^{(s_{1})}h^{(s_{2})}D^{(s_{3}-1)} \Big]
\end{eqnarray}
The detailed proof of this formula can be found in the Appendix where we describe also derivations of all other terms.

The next $O(D^{2})$ and $O(D^{3})$ level Lagrangians are
\begin{eqnarray}
 \mathcal{L}_{I}^{(2,0)}&=&\sum_{n_{i}} C_{n_1,n_2,n_3}^{s_1,s_2,s_3}\int dzdz_{1}dz_{2}dz_{3} \delta(z_{1}-z)\delta (z_{3}-z)\delta(z_{2}-z)\nonumber\\
  \Big[&+&\frac{s_{3}n_{3}s_{1}n_{1}}{2}\hat{T}(Q_{ij}|n_{1}-1,n_{2},n_{3}-1)D^{(s_{1}-1)}h^{(s_{2})}D^{(s_{3}-1)}\nonumber\\
  &+&\frac{s_{1}n_{1}s_{2}n_{2}}{2}\hat{T}(Q_{ij}|n_{1}-1,n_{2}-1,n_{3})D^{(s_{1}-1)}D^{(s_{2}-1)}h^{(s_{3})}\nonumber\\
  &+&\frac{s_{2}n_{2}s_{3}n_{3}}{2}\hat{T}(Q_{ij}|n_{1},n_{2}-1,n_{3}-1)h^{(s_{1})}D^{(s_{2}-1)}D^{(s_{3}-1)} \Big]\quad\quad
\end{eqnarray}
and
\begin{eqnarray}
  \mathcal{L}_{I}^{(3,0)}&=&\sum_{n_{i}} C_{n_1,n_2,n_3}^{s_1,s_2,s_3} \int dzdz_{1}dz_{2}dz_{3}  \delta(z_{1}-z)\delta (z_{3}-z)\delta(z_{2}-z)\nonumber\\
  \Big[&+&\frac{s_{3}n_{3}s_{2}n_{2}s_{1}n_{1}}{2}\hat{T}(Q_{ij}|n_{1}-1,n_{2}-1,n_{3}-1)D^{(s_{1}-1)}D^{(s_{2}-1)}D^{(s_{3}-1)} \Big]\nonumber\\
\end{eqnarray}
The remaining terms in the Lagrangian contain at least one trace:
\begin{eqnarray}
\mathcal{L}_{I}^{(0,1)}&=&\mathcal{L}_{I}^{(0,2)}=0\label{L01}\\
\mathcal{L}_{I}^{(0,3)}&=&\sum_{n_{i}} C_{n_{1},n_{2},n_{3}}^{s_{1},s_{2},s_{3}}\frac{Q_{12}Q_{23}Q_{31}}{8}\int dz_{1}dz_{2}dz_{3} \delta(z_{1}-z)\delta (z_{2}-z)\delta(z_{3}-z)\nonumber\\
  &&\Big[\hat{T}(Q_{12}-1,Q_{23}-1,Q_{31}-1|n_{1},n_{2},n_{3})\Box_{a}h^{(s_{1})}\Box_{b}h^{(s_{2})}\Box_{c}h^{(s_{3})}\Big]\quad\quad
\end{eqnarray}
\begin{eqnarray}
 \mathcal{L}_{I}^{(1,1)}&=&\sum_{n_{i}} C_{n_{1},n_{2},n_{3}}^{s_{1},s_{2},s_{3}}\int dz_{1}dz_{2}dz_{3} \delta(z_{1}-z)\delta (z_{2}-z)\delta(z_{3}-z)\nonumber\\
  \Big[&+&\frac{s_{1}Q_{12}n_{2}}{4}\hat{T}(Q_{12}-1,Q_{23},Q_{31}|n_{1},n_{2}-1,n_{3})
  D^{(s_{1}-1)}\Box_{b}h^{(s_{2})}h^{(s_{3})}\nonumber\\
  &+&\frac{s_{2}Q_{23}n_{3}}{4}\hat{T}(Q_{12},Q_{23}-1,Q_{31}|n_{1},n_{2},n_{3}-1)
  h^{(s_{1})}D^{(s_{2}-1)}\Box_{c}h^{(s_{3})}\nonumber\\
  &+&\frac{s_{3}Q_{31}n_{1}}{4}\hat{T}(Q_{12},Q_{23},Q_{31}-1|n_{1}-1,n_{2},n_{3})
  \Box_{a}h^{(s_{1})}h^{(s_{2})}D^{(s_{3}-1)} \Big]\quad\quad
\end{eqnarray}
\begin{eqnarray}
 \mathcal{L}_{I}^{(1,2)}&=&\sum_{n_{i}} C_{n_{1},n_{2},n_{3}}^{s_{1},s_{2},s_{3}}\int dz_{1}dz_{2}dz_{3} \delta(z_{1}-z)\delta (z_{2}-z)\delta(z_{3}-z)\nonumber\\
  \Big[&+&\frac{s_{1}Q_{12}Q_{23}n_{3}}{8}\hat{T}(Q_{12}-1,Q_{23}-1,Q_{31}|n_{1},n_{2},n_{3}-1)
  D^{(s_{1}-1)}\Box_{b}h^{(s_{2})}\Box_{c}h^{(s_{3})}\nonumber\\
  &+&\frac{s_{2}Q_{23}Q_{31}n_{1}}{8}\hat{T}(Q_{12},Q_{23}-1,Q_{31}-1|n_{1}-1,n_{2},n_{3})
  \Box_{a}h^{(s_{1})}D^{(s_{2}-1)}\Box_{c}h^{(s_{3})}\nonumber\\
  &+&\frac{s_{3}Q_{31}Q_{12}n_{2}}{8}\hat{T}(Q_{12}-1,Q_{23},Q_{31}-1|n_{1},n_{2}-1,n_{3})
  \Box_{a}h^{(s_{1})}\Box_{b}h^{(s_{2})}D^{(s_{3}-1)} \Big]\nonumber\\
\end{eqnarray}
\begin{eqnarray}
 \mathcal{L}_{I}^{(2,1)}&=&\sum_{n_{i}} C_{n_{1},n_{2},n_{3}}^{s_{1},s_{2},s_{3}}\int dz_{1}dz_{2}dz_{3} \delta(z_{1}-z)\delta (z_{2}-z)\delta(z_{3}-z)\nonumber\\
  \Big[&+&\frac{s_{2}s_{3}Q_{31}n_{1}n_{2}}{4}\hat{T}(Q_{12},Q_{23},Q_{31}-1|n_{1}-1,n_{2}-1,n_{3})
  \Box_{a}h^{(s_{1})}D^{(s_{2}-1)}D^{(s_{3}-1)}\nonumber\\
  &+&\frac{s_{1}s_{2}Q_{23}n_{3}n_{1}}{4}\hat{T}(Q_{12},Q_{23}-1,Q_{31}|n_{1}-1,n_{2},n_{3}-1)
  D^{(s_{1}-1)}D^{(s_{2}-1)}\Box_{c}h^{(s_{3})}\nonumber\\
  &+&\frac{s_{3}s_{1}Q_{12}n_{2}n_{3}}{4}\hat{T}(Q_{12}-1,Q_{23},Q_{31}|n_{1},n_{2}-1,n_{3}-1)
  D^{(s_{1}-1)}\Box_{b}h^{(s_{2})}D^{(s_{3}-1)} \Big]\nonumber\\
\end{eqnarray}
So we derived all cells of the following classification table corresponding to (\ref{intlag})
\begin{eqnarray}
\setlength{\unitlength}{0.254mm}
\begin{picture}(280,275)(125,-355)
        \allinethickness{0.254mm}\path(405,-80)(405,-175) 
        \allinethickness{0.254mm}\path(345,-235)(345,-80) 
        \allinethickness{0.254mm}\path(285,-295)(285,-80) 
        \allinethickness{0.254mm}\path(225,-355)(225,-80) 
        \allinethickness{0.254mm}\path(125,-80)(125,-355) 
        \allinethickness{0.254mm}\path(165,-80)(165,-355) 
        \allinethickness{0.254mm}\path(405,-80)(125,-80) 
        \allinethickness{0.254mm}\path(405,-175)(125,-175) 
        \allinethickness{0.254mm}\path(345,-235)(125,-235) 
        \allinethickness{0.254mm}\path(285,-295)(125,-295) 
        \allinethickness{0.254mm}\path(225,-355)(125,-355) 
        \allinethickness{0.254mm}\path(125,-115)(405,-115) 
        \allinethickness{0.254mm}\path(125,-80)(165,-115) 
        \put(130,-111){\shortstack{$\bar{h}$}} 
        \put(150,-96){\shortstack{$D$}} 
        \put(190,-106){\shortstack{$0$}} 
        \put(250,-106){\shortstack{$1$}} 
        \put(310,-106){\shortstack{$2$}} 
        \put(370,-106){\shortstack{$3$}} 
        \put(140,-151){\shortstack{$0$}} 
        \put(140,-211){\shortstack{$1$}} 
        \put(140,-271){\shortstack{$2$}} 
        \put(140,-331){\shortstack{$3$}} 
        \put(185,-151){\shortstack{$hhh$}} 
        \put(235,-151){\shortstack{$Dhh$}} 
        \put(295,-151){\shortstack{$DDh$}} 
        \put(355,-151){\shortstack{$DDD$}} 
        \put(185,-211){\shortstack{$0$}} 
        \put(185,-271){\shortstack{$0$}} 
        \put(185,-331){\shortstack{$\bar{h}\bar{h}\bar{h}$}} 
        \put(235,-211){\shortstack{$\bar{h}Dh$}} 
        \put(295,-211){\shortstack{$\bar{h}DD$}} 
        \put(235,-271){\shortstack{$\bar{h}\bar{h}D$}} 
\end{picture}
\end{eqnarray}
 and proved that after fixing the freedom of partial integration in the leading term (i.e. our cyclic ansatz)  all terms of interaction are determined in a unique way. More precisely we proved that \emph{for a given number of derivatives in the allowed range the cubic interaction of any three high spin fields is unique up to partial integration and field redefinition, and derived all these interactions}\footnote{The interactions discussed in this paper are all possible parity preserving interactions of bosonic fields. In some cases one should introduce charges to avoid triviality of the interaction, but the form of interaction is the same also for odd spin fields.}.

 Summarizing we see that the interaction Lagrangian in deDonder gauge ($D^{(s-1)}(z;a)=0$ for all three fields) can be expressed as a sum
\begin{eqnarray}
\mathcal{L}^{\emph{dD}}_{I}(h^{(s)})=\sum_{j=0}^{3} \mathcal{L}^{(0,j)}_{I}(h^{(s)}) .
\end{eqnarray}
and it is nothing else than the first column of this table. Therefore
(\ref{L01}) means that \emph{in deDonder gauge the traces of the HS fields can decouple from the fields as in the free Lagrangian.}

\section{Classification of gauge invariant cubic vertices}
\setcounter{equation}{0}
\quad From the recent literature \cite{MMR0}, \cite{M}, \cite{boulanger} we can first speculatively group the cubic gauge invariant
vertices into four classes depending on the number of Weyl tensors they contain. In \cite{boulanger} vertices with (say even) spins $s_1,s_2,s_3$
were explicitly constructed e.g. for the case $4,4,2$, and the result was presented in a form which was linear in the (linearized) Weyl tensor of the minimal spin. In the remaining bracket of terms a second Weyl tensor (say for spin 2) is apparently not hidden. In \cite{M} the ensemble of vertices constructed contain those of the spin type $2s,s,s$, and these are quadratic in the (linearized) Weyl tensor of the spin $s$ field. In all these cases the number of derivatives is minimal and equals $\Delta = s_3+s_2-s_1$, where $s_1$ is the minimal spin. In this section we order the spins  such as $s_1\leq s_2 \leq s_3$, because the smaller spins give rise to the
Weyl tensors, and these are represented by differential operators acting from left to right.
Abandoning the constraint of minimality on the number of derivatives, one can construct vertices from three Riemann tensors by contraction having $ \Delta_{BI} =s_1+s_2+s_3$ derivatives (Born-Infeld interaction). Finally a selfinteracting vertex of spin type $s, s, s$ has been derived in \cite{MMR0} for $s=4$, which apparently did not factorize into any Weyl tensor at all. If we consider only the leading order terms discussed in Section 2, the Weyl tensor reduces to the Riemann tensor. The Riemann tensor for spin $s$ gauge field \cite{DF}, \cite{MR6} is best defined for our purpose by the differential expression
\begin{equation}
R^{(s)}(z; a, b) = [(a\nabla)(b\partial_{c}) - (a\partial_{c})(b\nabla)]^{s} h^{(s)}(z; c)\label{4.1}
\end{equation}
With the cyclic ansatz for a vertex of an arbitrary spin type presented in the preceeding section, we can derive results on the factorization of the l. o. terms into Riemann tensors. Denoting the powers of the Riemann tensors maximally appearing by $n$,
and letting the number of derivatives be $\Delta* \geq \Delta_{min}$ we may define classes of vertices $\mathcal{V}_{n}(\Delta*)$ and characterize them in terms of the spin $s_{i}$. There remain two tasks.
First we may consider all explicitly known vertices for minimal derivative number $\Delta_{min}$ and rewrite their leading orders in terms of the cyclic notation for the l. o. terms discussed in the preceeding section. This was in all cases possible. On the other hand we can turn this issue around and derive from the cyclic ansatz of Section 1 with minimal derivative number the maximum amount of information possible on the factorization into Riemann tensors, which gives new general insights on which spin type belongs to which class $\mathcal{V}_{n}(\Delta_{min})$.

We prove the following theorem: \emph{The class $\mathcal{V}_0(\Delta_{min})$ consists of those spin types with $s_3 < 2 s_1$, the class $\mathcal{V}_{1}(\Delta_{min})$  of the spin types $s_3 \geq 2 s_1, s_2> s_1 $, and the class $\mathcal{V}_2(\Delta_{min})$ of the types $s_3 \geq 2 s_1 = 2 s_2$ }. As a corollary we obtain that $\mathcal{V}_0(\Delta_{min})$ contains all selfinteractions.

For the proof we start from (\ref{delta}), (\ref{2.25}). We obtain by replacing $n_1 = s_1-p$ and summing over $n_2,n_3$
\begin{eqnarray}
\sum_{n_2,n_3}{ s_1 \choose n_1,n_2-s_2+s_1,n_3-s_3+s_1}(\partial_{a}\nabla_2)^{n_1}(\partial_{b}\nabla_3)^{n_2}(\partial_{c}
\nabla_1)^{n_3}\nonumber\\
(\partial_{a}\partial_{b})^{n_3-s_3+s_1} (\partial_{b}\partial_{c})^{n_1}(\partial_{c}\partial_{a})^{n_2-s_2+s_1}
h^{(s_1)}(z_1;a)h^{(s_2)}(z_2;b)h^{(s_3)}(z_3;c)\nonumber\\
={s_1 \choose p} [(\partial_{a}\nabla_2)(\partial_{b}\partial_{c})]^{s_1-p}(\partial_{c}\nabla_1)^{s_3-s_1}(\partial_{b}\nabla_3)^{s_2-s_1}\nonumber\\
\{ (\partial_{c}\nabla_1)(\partial_{b}\partial_{a}) + (\partial_{c}\partial_{a})(\partial_{b}\nabla_3) \}^{p}\quad   h^{(s_1)}(z_1;a)h^{(s_2)}(z_2;b) h^{(s_3)}(z_3;c)\label{4.2}
\end{eqnarray}
Since we want to neglect all higher order terms  (such as divergences, traces and contractions of gradients $(\nabla_{i}\nabla_{j})$\footnote{Note that $(\nabla_{i}\nabla_{j})$ product of derivatives in this case leads also to the $\Box R^{(s)}(a;b;h^{(s)}(c;z))$ which is on shell equal to $R^{(s)}(a;b;(c\nabla)D^{(s-1)})=0$ due to gauge invariance of the linearized curvature}), use partial integrations implied by $\nabla_1+\nabla_2+\nabla_3 = 0$, and keep only the l.o. expressions, we can in the square bracket replace $(\partial_{b}\nabla_3)$ by $-(\partial_{b}\nabla_1)$. Introducing two arbitrary tangential vectors $e,f$ and the star product we get
\begin{equation}
[(\partial_{b}\partial_{a})(\partial_{c}\nabla_1)-(\partial_{c}\partial_{a})(\partial_{b}\nabla_1)]^{p} =
[(e\partial_{a})(f\nabla_1) - (f\partial_{a})(e\nabla_1]^{p} *_{e}*_{f} [(e\partial_{b})(f\partial_{c})]^{p}\label{4.3}
\end{equation}
where we obtained $p$ differential operator factors acting on $h^{(s_1)}(z_1,a)$ of a Riemann tensor. The other $s_1-p$ components are produced as follows.

The remaining factors in (\ref{4.2}) can be reordered by
\begin{eqnarray}
&&[(\partial_{a}\nabla_2)(\partial_{b}\partial_{c})]^{s_1-p}(\partial_{c}\nabla_1)^{s_3-s_1}(\partial_{b}\nabla_3)^{s_2-s_1}\nonumber\\
&&=[(\partial_{a}\nabla_2)(\partial_{c}\nabla_1)]^{s_1-p}(\partial_{b}
\partial_{c})^{s_1-p}(\partial_{c}\nabla_1)^{s_3-2s_1+p}(\partial_{b}\nabla_3)^{s_2-s_1}\label{4.4}
\end{eqnarray}
which is permitted if $s_3-2s_1 \geq 0$, since $p$ runs from $0$ to $s_1$.
Then the square bracket can be contracted in the same way with the tangential vectors $e,f$, and neglecting a contraction $(\nabla_1\nabla_2)$ as of higher order, we get
\begin{equation}
[(e\partial_{a})(f\nabla_1)-(f\partial_{a})(e\nabla_1)]^{s_1-p} *_{e}*_{f}[(e \nabla_2)(f \partial_{c})]^{s_1-p}\label{4.5}
\end{equation}
Multiplication of the two products over the square bracket operators (\ref{4.3}) and (\ref{4.5}) yields the Riemann tensor
\begin{equation}
R^{s_1}(z_1;f,e) = [(e\partial_{a})(f\nabla_1)-(f\partial_{a})(e\nabla_1)]^{s_1} h^{(s_1)}(z_1;a)\label{4.6}
\end{equation}
However, in the remaining factors there is still the term
\begin{equation}
(\partial_{c}\nabla_1)^{s_3-2s_1+p}\label{4.7}
\end{equation}
which acts on $h^{(s_1)}$ (i.e. the Riemann tensor), but by partial integration and neglect of divergences as higher order terms, we can let it act on $h^{(s_2)}$. Then we obtain the first Riemann tensor formula
\begin{eqnarray}
&&R^{(s_1)}(z_1;f,e)*_{e}*_{f}\sum_{p= 0}^{s_1}(-1)^{p}{s_1 \choose p}(e\partial_{b})^{p}(e\nabla_2)^{s_1-p}(f\partial_{c})^{s_1}
\nonumber\\
&& (\partial_{b}\partial_{c})^{s_1-p}(\partial_{c}\nabla_2)^{s_3-2s_1+p}(\partial_{b}\nabla_3)^{s_2-s_1} \quad h^{(s_2)}(z_2;b) h^{(s_3)}(z_3;c)\label{4.8}
\end{eqnarray}

The next construction is also intended to be a Riemann tensor for the spin $s_1$. For this purpose we construct a homogeneous function in $b$ of degree $s_1$. Using the appropriate factors from (\ref{4.8}) we define
\begin{equation}
H(z_2,z_3;b,c,f) = (f\partial_{c})^{s_1}(\partial_b\nabla_3)^{s_2-s_1}(\partial_{c}\nabla_2)^{s_3-2s_1} h^{(s_2)}(z_2;b)h^{(s_3)}(z_3;c)\label{4.9}
\end{equation}
which is homogeneous both in $b$ and $c$ of degree $s_1$. Then we can perform the sum in (\ref{4.8}) and get
\begin{equation}
[(e\nabla_2)(\partial_{b}\partial_{c}) - (\partial_{c}\nabla_2)(e\partial_{b})]^{s_1}H(z_2,z_3;b,c,f)\label{4.10}
\end{equation}
Concerning the differential operator in front of $H$ this expression looks like a Riemann tensor $R^{(s_1)}(z_2)$.
But a gauge transformation $\delta   h^{(s_2)}(z_2,b)$ acts as
\begin{eqnarray}
(\partial_{b}\nabla_3)^{s_2-s_1} (b\nabla_2) \epsilon^{(s_2-1)}(z_2;b)= \nonumber\\
(s_2-s_1)(\nabla_2\nabla_3)(\partial_{b}\nabla_3)^{s_2-s_1-1}\epsilon ^{(s_2-1)}(z_2;b)\nonumber\\  +(b \nabla_2)
(\partial_{b}\nabla_3)^{s_2-s_1}
\epsilon^{(s_2-1)}(z_2,b)\label{4.11}
\end{eqnarray}
We conclude that only for $s_2= s_1$ the expression (\ref{4.10}) can be considered as a proper Riemann tensor. This completes the proof of the theorem.

\section{Discussion: Towards gauge transformations as Lie algebras}
\setcounter{equation}{0}
We can derive the first order gauge transformation of $h^{(s_1)}$
from the r. h. s. of Noether's equation (\ref{2.28}) taken off shell
\begin{eqnarray}
&&[O(\mathcal{F}) \quad\textnormal{part of}\quad\delta^{0}_{\epsilon^{(s_{1}-1)}} \mathcal{L}_{I}]\nonumber\\
&=&\sum_{n_{i}} C_{n_{1},n_{2},n_{3}}^{s_{1},s_{2},s_{3}}\int dz_{1}dz_{2}dz_{3} \delta(z_{1}-z)\delta (z_{2}-z)\delta(z_{3}-z)\nonumber\\
  \Big[&-&\frac{s_{1}n_{1}}{2}\hat{T}(Q_{ij}|n_{1}-1,n_{2},n_{3})
  \epsilon^{(s_{1}-1)}\mathcal{F}^{(s_{2})}h^{(s_{3})}\nonumber\\
  &+&\frac{s_{1}Q_{12}n_{2}}{4}\hat{T}(Q_{12}-1,Q_{23},Q_{31}|n_{1},n_{2}-1,n_{3})
  \epsilon^{(s_{1}-1)}\Box_{b}\mathcal{F}^{(s_{2})}h^{(s_{3})}\nonumber\\
  &+&\frac{s_{3}s_{1}Q_{12}n_{2}n_{3}}{4}\hat{T}(Q_{12}-1,Q_{23},Q_{31}|n_{1},n_{2}-1,n_{3}-1)
  \epsilon^{(s_{1}-1)}\Box_{b}\mathcal{F}^{(s_{2})}D^{(s_{3}-1)} \nonumber\\
  &+&\frac{s_{1}Q_{12}Q_{23}n_{3}}{8}\hat{T}(Q_{12}-1,Q_{23}-1,Q_{31}|n_{1},n_{2},n_{3}-1)
  \epsilon^{(s_{1}-1)}\Box_{b}\mathcal{F}^{(s_{2})}\Box_{c}h^{(s_{3})}\nonumber\\
  &+&\frac{s_{1}n_{1}}{2}\hat{T}(Q_{ij}|n_{1}-1,n_{2},n_{3})
  \epsilon^{(s_{1}-1)}h^{(s_{2})}\mathcal{F}^{(s_{3})}\nonumber\\
  &+&\frac{s_{1}Q_{12}n_{2}}{4}\hat{T}(Q_{12}-1,Q_{23},Q_{31}|n_{1},n_{2}-1,n_{3})
  \epsilon^{(s_{1}-1)}\Box_{b}h^{(s_{2})}\mathcal{F}^{(s_{3})}\nonumber\\
  &+&\frac{s_{1}s_{2}n_{1}n_{2}}{4}\hat{T}(Q_{12},Q_{23},Q_{31}|n_{1}-1,n_{2}-1,n_{3})
  \epsilon^{(s_{1}-1)}D^{(s_{2}-1)}\mathcal{F}^{(s_{3})} \nonumber\\
  &-&\frac{s_{1}Q_{12}Q_{23}n_{3}}{8}\hat{T}(Q_{12}-1,Q_{23}-1,Q_{31}|n_{1},n_{2},n_{3}-1)
  \epsilon^{(s_{1}-1)}\Box_{b}h^{(s_{2})}\Box_{c}\mathcal{F}^{(s_{3})}\Big]\nonumber\\\label{5.1}
\end{eqnarray}
  If we assume moreover that the gauge transformations form a Lie algebra of power series in some "coupling constant"
$g$, we can following along the ideas of Berends, Burger and Van Dam in their classical paper \cite{vanDam2}
derive conclusions on the higher order interactions. We sum up simple results:

(1) The arguments of these authors to show that such
power series algebra does not exist for $s=3$, cannot be generalized to even spins;

(2) the quartic interaction Lagrangian is nonzero and contains $2s-2$ derivatives.

For a given gauge function $\epsilon^{(s-1)}(z;a)$ the gauge transformation is a substitution (classically) with expansion
\begin{equation}
h \rightarrow h + \delta_{\epsilon}h = h + \nabla \epsilon + \sum_{n\geq 1} g^{n}\Theta_{n}(h,h,...h;\epsilon)\label{5.2}
\end{equation}
with $\Theta_{n}$ depending on $\epsilon$ linearly and on $h$ in the n'th power. Moreover we assume that the commutator of two such transformations is given by
\begin{equation}
[\delta_{\epsilon},\delta_{\eta}]h  = \delta_{C(h;\epsilon,\eta)} h \label{5.3}
\end{equation}
with the expansion
\begin{equation}
C(h;\epsilon,\eta) = g\sum_{n\geq 0} g^{n}C_{n}(h, h,...h;\epsilon,\eta) \label{5.4}
\end{equation}
where each $C_{n}$ depends on $\epsilon$ and $\eta$ linearly and on $h$ in the $n$'th power. As substitutions gauge transformations are associative and their infinitesimals must satisfy the Jacobi identity. At order $g^{2}$ this is e.g.
\begin{equation}
\sum_{\eta,\epsilon,\zeta \textnormal{cyclic}}\{C_1(\nabla\zeta;\eta,\epsilon) +C_0(C_0(\eta,\epsilon),\zeta)\} = 0 \label{5.5}
\end{equation}
The commutator can also be expanded
\begin{eqnarray}
[\delta_{\eta},\delta_{\epsilon}] &=& g(\Theta_1(\nabla\epsilon;\eta) -\Theta_1(\nabla\eta;\epsilon))\nonumber\\
&+& g^{2}\{[\Theta_1(\Theta_1(h;\epsilon);\eta) -\Theta_1(\Theta_1(h;\eta);\epsilon)]\nonumber\\
&+& [\Theta_2(\nabla\epsilon,h;\eta) -\Theta_2(\nabla\eta,h;\epsilon)]\nonumber\\
&+& [\Theta_2(h,\nabla\epsilon;\eta) -\Theta_2(h,\nabla\eta;\epsilon)]\}
+O(g^{3}) \label{5.6}
\end{eqnarray}
Inserting this expansion into the definition of the functions $C_{n}$ we obtain
\begin{eqnarray}
\nabla C_0(\eta,\epsilon) &=& \Theta_1(\nabla\epsilon;\eta) - \Theta_1(\nabla\eta;\epsilon)  \label{5.7}\\
\nabla C_1(h; \eta,\epsilon) &=& \Theta_1(\Theta_1(h;\epsilon);\eta) -\Theta_1(\Theta_1(h;\eta);\epsilon)
 - \Theta_1(h; C_0(\eta,\epsilon)) \nonumber\\
 &+& \Theta_2(\nabla\epsilon,h;\eta) -\Theta_2(\nabla\eta,h;\epsilon) +\Theta_2(h, \nabla\epsilon; \eta) - \Theta_2(h, \nabla\eta;\epsilon)\label{5.8}\nonumber\\
\end{eqnarray}
Assume that $\Theta_1(h;\epsilon)$ has been extracted from (\ref{5.1}) for the case of equal spins $s$ and the minimal number of derivatives. Then the order of derivations in $\Theta_1$ is
\begin{equation}
\natural\Theta_1(h; \epsilon) = s-1 \label{5.9}
\end{equation}
Inserting this result into (\ref{5.5}), (\ref{5.7}) we obtain the number of derivations in $C_0, C_1$ as
\begin{equation}
\natural C_0(\eta,\epsilon) = s-1 \quad\textnormal{and}\quad \natural C_1(h;\eta,\epsilon) = 2s-3
\end{equation}
This implies
\begin{equation}
\natural\Theta_2(h,h;\epsilon) = 2s-3
\end{equation}
Consequently the quartic interaction must contain $2s-2$ derivatives. The argument can be continued to still higher interactions: For n'th order interactions the number is $(s-2)(n-2)+2$. This result is equivalent to introducing a scale $L$ and dimensions
in the following way
\begin{equation}
[h] = L^{s-2}, \quad [\nabla] = \frac{1}{L}
\end{equation}
with a dimensionless coupling constant $g$, so that each term in the power series has the same dimension. Note that in the case of $\Delta=\Delta_{min}$ we obtained in the previous sections and in \cite{MMR0} the same dimensions for cubic selfinteractions  and free part of Fronsdal action.

In \cite{vanDam2} the argument was presented that for spin $s=3$ a Lie algebra of gauge transformations in the form of power series
does not exist, the problem starting with the second power.  The argument was based on the term
\begin{equation}
(\partial_{a}\nabla_2)^{s-1}\epsilon^{(s-1)}(z_1;a) h^{(s)}(z_2;b)
\end{equation}
which exists in $\Theta_1$. Such term is present in fact for any spin, as can be inspected from (\ref{5.1}). Namely, in the fifth term of the square bracket of (\ref{5.1}) (this is the unique localization) we get such expression for $n_1=s,n_2=n_3=0$. In equation (\ref{5.8}) in the first line we have thus $2s-2$ derivations acting on the field $h$ in either term. In no other terms of (\ref{5.8}) such expression appears. Therefore they must cancel inside this line and they do cancel indeed for even spin only. There is in this case no obstruction of the power series algebra by these arguments. A deeper investigation of such algebras will follow in the future.

\subsection*{Acknowledgements}
This work is supported in part by Alexander von Humboldt Foundation under 3.4-Fokoop-ARM/1059429.
Work of K.M. was made with partial support of CRDF-NFSAT UCEP06/07 and CRDF-NFSAT-SCS MES RA ECSP 09\_01.

\section*{Appendix \\ Proof of  $\mathcal{L}_{I}^{(i,j)}$}
\setcounter{equation}{0}
\renewcommand{\theequation}{A.\arabic{equation}}
\quad
The expression for $\mathcal{L}_{I}^{(1,0)}$ is right if the following remaining group of terms vanishes:
\begin{eqnarray}
  && +\frac{s_{1}n_{1}s_{3}}{2}C_{(n_i)}^{(s_i)}[\hat{T}(Q_{ij}|n_{1}-1,n_{2},n_{3}),(c\nabla_{3})]
  (\epsilon^{(s_{1}-1)}h^{(s_{2})}D^{(s_{3}-1)}+D^{(s_{1}-1)}h^{(s_{2})}\epsilon^{(s_{3}-1)})\qquad\quad\label{2.38}\\
  && +\frac{s_{1}n_{1}s_{2}}{2}C_{(n_i)}^{(s_i)}[\hat{T}(Q_{ij}|n_{1}-1,n_{2},n_{3}),(b\nabla_{2})]
  (D^{(s_{1}-1)}\epsilon^{(s_{2}-1)}h^{(s_{3})}-\epsilon^{(s_{1}-1)}D^{(s_{2}-1)}h^{(s_{3})})\qquad\quad \label{2.39}\\
  && +\frac{s_{2}n_{2}s_{1}}{2}C_{(n_i)}^{(s_i)}[\hat{T}(Q_{ij}|n_{1},n_{2}-1,n_{3}),(a\nabla_{1})]
  (D^{(s_{1}-1)}\epsilon^{(s_{2}-1)}h^{(s_{3})}+\epsilon^{(s_{1}-1)}D^{(s_{2}-1)}h^{(s_{3})})\qquad\quad \label{2.40}\\
  &&+\frac{s_{2}n_{2}s_{3}}{2}C_{(n_i)}^{(s_i)}[\hat{T}(Q_{ij}|n_{1},n_{2}-1,n_{3}),(c\nabla_{3})]
  (h^{(s_{1})}D^{(s_{2}-1)}\epsilon^{(s_{3}-1)}-h^{(s_{1})}\epsilon^{(s_{2}-1)}D^{(s_{3}-1)})\qquad\quad \label{2.41}\\
  && +\frac{s_{3}n_{3}s_{2}}{2}C_{(n_i)}^{(s_i)}[\hat{T}(Q_{ij}|n_{1},n_{2},n_{3}-1),(b\nabla_{2})]
  (h^{(s_{1})}D^{(s_{2}-1)}\epsilon^{(s_{3}-1)}+h^{(s_{1})}\epsilon^{(s_{2}-1)}D^{(s_{3}-1)})\qquad\quad  \label{2.42}\\
  && +\frac{s_{3}n_{3}s_{1}}{2}C_{(n_i)}^{(s_i)}[\hat{T}(Q_{ij}|n_{1},n_{2},n_{3}-1),(a\nabla_{1})]
  (\epsilon^{(s_{1}-1)}h^{(s_{2})}D^{(s_{3}-1)}-D^{(s_{1}-1)}h^{(s_{2})}\epsilon^{(s_{3}-1)})\qquad\quad  \label{2.43}\\
  && - s_{1}Q_{12}s_{2}C_{(n_i)}^{(s_i)}\hat{T}(Q_{12}-1,Q_{23},Q_{31}|n_{i})\epsilon^{(s_{1}-1)}D^{(s_{2}-1)}h^{(s_{3})} \label{2.44}\\
  && - s_{2}Q_{23}s_{3}C_{(n_i)}^{(s_i)}\hat{T}(Q_{12},Q_{23}-1,Q_{31}|n_{i})h^{(s_{1})}\epsilon^{(s_{2}-1)}D^{(s_{3}-1)} \label{2.45}\\
  && - s_{3}Q_{31}s_{1}C_{(n_i)}^{(s_i)}\hat{T}(Q_{12},Q_{23},Q_{31}-1|n_{i})D^{(s_{1}-1)}h^{(s_{2})}\epsilon^{(s_{3}-1)} \label{2.46}
\end{eqnarray}
Indeed calculating commutators in the leading order and using relation  (\ref{2.34}) we see that
\begin{eqnarray}
  && (\ref{2.38})+(\ref{2.43})\nonumber \\
  && =s_{1}s_{2}(Q_{23}+1)C^{s_{i}}_{n_{1}+1,n_{2},n_{3}}\hat{T}(Q_{12},Q_{23},Q_{31}|n_{1},n_{2}+1,n_{3})
  D^{(s_{1}-1)}h^{(s_{2})}\epsilon^{(s_{3}-1)} \quad\qquad
\end{eqnarray}
which exactly cancels (\ref{2.46}) after a corresponding shift of $n_{2}$ and using relation (\ref{2.35}). In a similar way we can prove cancellation of the other two sets of three lines.

To prove formulas for $\mathcal{L}_{I}^{(2,0)}$ and $\mathcal{L}_{I}^{(3,0)}$  we should manage the commutators of $T$ operators with $a,b,c$, gradients in the following expression
\begin{eqnarray}
&&\frac{s_{1}s_{2}s_{3}}{2} C_{(n_i)}^{(s_i)}\Big[\nonumber\\
  &&[n_{1}n_{3}\hat{T}(Q_{ij}|n_{1}-1,n_{2},n_{3}-1),(b\nabla_{2})]
  (D^{(s_{1}-1)}D^{(s_{2}-1)}\epsilon^{(s_{3}-1)}+D^{(s_{1}-1)}\epsilon^{(s_{2}-1)}D^{(s_{3}-1)})\qquad\quad\nonumber\\
  && [n_{2}n_{3}\hat{T}(Q_{ij}|n_{1},n_{2}-1,n_{3}-1),(a\nabla_{1})]
  (D^{(s_{1}-1)}\epsilon^{(s_{2}-1)}D^{(s_{3}-1)}+\epsilon^{(s_{1}-1)}D^{(s_{2}-1)}D^{(s_{3}-1)})\qquad\quad\nonumber\\
  && [n_{1}n_{2}\hat{T}(Q_{ij}|n_{1}-1,n_{2}-1,n_{3}),(c\nabla_{3})]
  (\epsilon^{(s_{1}-1)}D^{(s_{2}-1)}D^{(s_{3}-1)}+D^{(s_{1}-1)}D^{(s_{2}-1)}\epsilon^{(s_{3}-1)})\qquad\quad\nonumber\\
  && - n_{3}Q_{12}\hat{T}(Q_{12}-1,Q_{23},Q_{31}|n_{1},n_{2},n_{3}-1)(\epsilon^{(s_{1}-1)}D^{(s_{2}-1)}D^{(s_{3}-1)}
  -\epsilon^{(s_{1}-1)}D^{(s_{2}-1)}D^{(s_{3}-1)})\nonumber\\
  && - n_{2}Q_{31}\hat{T}(Q_{12},Q_{23},Q_{31}-1|n_{1},n_{2}-1,n_{3})(D^{(s_{1}-1)}D^{(s_{2}-1)}\epsilon^{(s_{3}-1)}
  -D^{(s_{1}-1)}\epsilon^{(s_{2}-1)}D^{(s_{3}-1)})\nonumber\\
  && - n_{1}Q_{23}\hat{T}(Q_{12},Q_{23}-1,Q_{31}|n_{1}-1,n_{2},n_{3})(D^{(s_{1}-1)}
  \epsilon^{(s_{2}-1)}D^{(s_{3}-1)}-\epsilon^{(s_{1}-1)}D^{(s_{2}-1)}D^{(s_{3}-1)})\Big]\nonumber\\ \label{2.55}
\end{eqnarray}
and use again (\ref{2.33})-(\ref{2.35}) to show that (\ref{2.55}) is zero.

The remaining terms are:
\begin{eqnarray}
  &&\frac{1}{2} C_{(n_i)}^{(s_i)}\Big[\nonumber\\
  && -s_{1}Q_{12}s_{2}(s_{2}-1)[\hat{T}(Q_{12}-1,Q_{23},Q_{31}|n_{i}),(b\nabla_{2})]\epsilon^{(s_{1}-1)}\bar{h}^{(s_{2}-2)}h^{(s_{3})}\nonumber\\
  &&-s_{2}Q_{23}s_{3}(s_{3}-1)[\hat{T}(Q_{12},Q_{23}-1,Q_{31}|n_{i}),(c\nabla_{3})]h^{(s_{1})}\epsilon^{(s_{2}-1)}\bar{h}^{(s_{3}-2)}\nonumber\\
  && -s_{3}Q_{31}s_{1}(s_{1}-1)[\hat{T}(Q_{12},Q_{23},Q_{31}-1|n_{i}),(a\nabla_{1})]\bar{h}^{(s_{1}-2)}h^{(s_{2})}\epsilon^{(s_{3}-1)}\Big]
\end{eqnarray}
and
\begin{eqnarray}
&&\frac{s_{1}s_{2}s_{3}}{4} C_{(n_i)}^{(s_i)}\Big[ \nonumber\\
  && - n_{3}Q_{12}(s_{2}-1)[\hat{T}(Q_{12}-1,Q_{23},Q_{31}|n_{1},n_{2},n_{3}-1),(b\nabla_{2})]\nonumber\\
  && (\epsilon^{(s_{1}-1)} \bar{h}^{(s_{2}-2)}D^{(s_{3}-1)}-D^{(s_{1}-1)}\bar{h}^{(s_{2}-2)}\epsilon^{(s_{3}-1)})\nonumber\\
  &&- n_{1}Q_{23}(s_{3}-1)[\hat{T}(Q_{12},Q_{23}-1,Q_{31}|n_{1}-1,n_{2},n_{3}),(c\nabla_{3})]\nonumber\\
  &&(D^{(s_{1}-1)}\epsilon^{(s_{2}-1)}\bar{h}^{(s_{3}-2)}
  -\epsilon^{(s_{1}-1)}D^{(s_{2}-1)}\bar{h}^{(s_{3}-2)}) \nonumber\\
  &&- n_{2}Q_{31}(s_{1}-1)[\hat{T}(Q_{12},Q_{23},Q_{31}-1|n_{1},n_{2}-1,n_{3}),(a\nabla_{1})]\nonumber\\
  &&(\bar{h}^{(s_{1}-2)}D^{(s_{2}-1)}\epsilon^{(s_{3}-1)}
  -\bar{h}^{(s_{1}-2)}\epsilon^{(s_{2}-1)}D^{(s_{3}-1)})\Big] \label{2.60}
 \end{eqnarray}
 \begin{eqnarray}
  && -\frac{s_{3}s_{1}(s_{1}-1)n_{1}}{4}C_{(n_i)}^{(s_i)}Q_{31}\hat{T}(Q_{12},Q_{23},Q_{31}-1|n_{1}-1,n_{2},n_{3})\nonumber\\
  &&(\delta\bar{h}^{(s_{1}-2)}h^{(s_{2})}D^{(s_{3}-1)} +2 (\nabla D)^{(s_{1}-2)}h^{(s_{2})}\epsilon^{(s_{3}-1)}) \nonumber\\
  && -\frac{s_{1}s_{2}(s_{2}-1)n_{2}}{4}C_{(n_i)}^{(s_i)}Q_{12}\hat{T}(Q_{12}-1,Q_{23},Q_{31}|n_{1},n_{2}-1,n_{3})\nonumber\\
  &&(D^{(s_{1}-1)}\delta\bar{h}^{(s_{2}-2)}h^{(s_{3})} +2\epsilon^{(s_{1}-1)}(\nabla D)^{(s_{2}-2)}h^{(s_{3})}) \nonumber\\
  && -\frac{s_{2}s_{3}(s_{3}-1)n_{3}}{4}C_{(n_i)}^{(s_i)}Q_{23}\hat{T}(Q_{12},Q_{23}-1,Q_{31}|n_{1},n_{2},n_{3}-1)\nonumber\\
  &&(h^{(s_{1})}D^{(s_{2}-1)}\delta\bar{h}^{(s_{3}-2)} +2h^{(s_{1})}\epsilon^{(s_{2}-1)}(\nabla D)^{(s_{3}-2)})
\end{eqnarray}
The last $DD\bar{h}$ terms coming from our calculation are:
\begin{eqnarray}
&&\frac{s_{1}s_{2}s_{3}}{4} C_{(n_i)}^{(s_i)}\Big[\nonumber\\
  && - (s_{3}-1)n_{1}n_{3}Q_{23}\hat{T}(Q_{12},Q_{23}-1,Q_{31}|n_{1}-1,n_{2},n_{3}-1)\nonumber\\
  &&(D^{(s_{1}-1)}D^{(s_{2}-1)}\delta\bar{h}^{(s_{3}-2)}
  +2D^{(s_{1}-1)}\epsilon^{(s_{2}-1)}(\nabla D)^{(s_{3}-2)})\nonumber\\
  &&- (s_{2}-1)n_{2}n_{3}Q_{12}\hat{T}(Q_{12}-1,Q_{23},Q_{31}|n_{1},n_{2}-1,n_{3}-1)\nonumber\\
  &&(D^{(s_{1}-1)}\delta\bar{h}^{(s_{2}-2)}D^{(s_{3}-1)}
  +2\epsilon^{(s_{1}-1)}(\nabla D)^{(s_{2}-2)}D^{(s_{3}-1)}) \nonumber\\
  &&- (s_{1}-1)n_{1}n_{2}Q_{31}\hat{T}(Q_{12},Q_{23},Q_{31}-1|n_{1}-1,n_{2}-1,n_{3})\nonumber\\
  &&(\delta\bar{h}^{(s_{1}-2)}D^{(s_{2}-1)}D^{(s_{3}-1)}
  +2(\nabla D)^{(s_{3}-2)}D^{(s_{2}-1)}\epsilon^{(s_{3}-1)})\Big] \label{2.69}
 \end{eqnarray}
These terms can be used in the same fashion to prove the remaining set of $\mathcal{L}_{I}^{(i,j)}, j \not= 0$ that contain traces.


\begin{thebibliography}{100}
\bibitem{MMR0}
  R.~Manvelyan, K.~Mkrtchyan and W.~R\"uhl,
  ``Direct construction of a cubic selfinteraction for higher spin gauge
  fields,'' Nucl.Phys. B {\bf 844} (2011) 348-364,
  arXiv:1002.1358 [hep-th].
\bibitem{M}
  R.~Manvelyan, K.~Mkrtchyan and W.~R\"uhl,
  ``Off-shell construction of some trilinear higher spin gauge field
  interactions,''
  Nucl.\ Phys.\  B {\bf 826} (2010) 1
  [arXiv:0903.0243 [hep-th]].
\bibitem{MMR}
R.~Manvelyan and K.~Mkrtchyan,
``Conformal invariant interaction of a scalar field with the higher spin
  field in $AdS_{D}$,'' Mod.Phys.Lett. A {\bf 25} (2010) 1333-1348 [arXiv:0903.0058 [hep-th]].
\bibitem{MR}
R.~Manvelyan and W.~R\"uhl, ``Conformal coupling of higher spin
gauge fields to a scalar field in AdS(4) and generalized Weyl
invariance,'' Phys.\ Lett.\ B {\bf 593} (2004) 253,
[arXiv:hep-th/0403241].
\bibitem{vanDam}
  F.~A.~Berends, G.~J.~H.~Burgers and H.~van Dam,
  ``Explicit Construction Of Conserved Currents For Massless Fields Of
  Arbitrary Spin,'' Nucl.\ Phys.\  B {\bf 271} (1986) 429;
\bibitem{vanDam1}
   F.~A.~Berends, G.~J.~H.~Burgers and H.~Van Dam,
  ``On Spin Three Selfinteractions,''
  Z.\ Phys.\  C {\bf 24} (1984) 247;
\bibitem{vanDam2}
  F.~A.~Berends, G.~J.~H.~Burgers and H.~van Dam,
  ``On The Theoretical Problems In Constructing Interactions Involving Higher
  Spin Massless Particles,''
  Nucl.\ Phys.\  B {\bf 260} (1985) 295.
\bibitem{Vasiliev}
  E.~S.~Fradkin and M.~A.~Vasiliev, ``On The Gravitational Interaction
  Of Massless Higher Spin Fields,'' Phys.\ Lett.\ B {\bf 189} (1987)
  89.
\bibitem{Vasiliev1}
   E.~S.~Fradkin and M.~A.~Vasiliev, ``Cubic Interaction In
  Extended Theories Of Massless Higher Spin Fields,'' Nucl.\ Phys.\ B
  {\bf 291} (1987) 141.
  M.~A.~Vasiliev, "Cubic Interactions of Bosonic Higher Spin Gauge Fields in $AdS_{5}$",
  [arXiv:hep-th/0106200].
  M.~A.~Vasiliev, "N = 1 Supersymmetric Theory of Higher Spin Gauge Fields in $AdS_{5}$ at the Cubic Level",
  [arXiv:hep-th/0206068]
\bibitem{Damour}
  T.~Damour and S.~Deser,
  ``Geometry of spin 3 gauge theories,''
  Annales Poincare Phys.\ Theor.\  {\bf 47}, 277 (1987);
 T.~Damour and S.~Deser,
  ``Higher derivative interactions of higher spin gauge fields,''
  Class.\ Quant.\ Grav.\  {\bf 4}, L95 (1987).
\bibitem{Metsaev}
  R.~R.~Metsaev,
  ``Cubic interaction vertices for massive and massless higher spin fields,''
  Nucl.\ Phys.\  B {\bf 759} (2006) 147
  [arXiv:hep-th/0512342];
\bibitem{Metsaev1}
  R.~R.~Metsaev,  ``Cubic interaction vertices for fermionic and bosonic arbitrary spin  fields,''
  arXiv:0712.3526 [hep-th].
\bibitem{Augusto}
 A.~Sagnotti,
  ``Higher Spins and Current Exchanges,''
  arXiv:1002.3388 [hep-th];
 D.~Francia, J.~Mourad and A.~Sagnotti,
  ``Current exchanges and unconstrained higher spins,''
  Nucl.\ Phys.\  B {\bf 773} (2007) 203
  [arXiv:hep-th/0701163];  D.~Francia and A.~Sagnotti,
  ``Higher-spin geometry and string theory,''
  J.\ Phys.\ Conf.\ Ser.\  {\bf 33} (2006) 57
  [arXiv:hep-th/0601199]; A.~Sagnotti and M.~Tsulaia,
  ``On higher spins and the tensionless limit of string theory,''
  Nucl.\ Phys.\  B {\bf 682} (2004) 83
  [arXiv:hep-th/0311257];
  D.~Francia and A.~Sagnotti,
  ``On the geometry of higher-spin gauge fields,''
  Class.\ Quant.\ Grav.\  {\bf 20} (2003) S473
  [arXiv:hep-th/0212185].
\bibitem{ouvry}
  I.~G.~Koh, S.~Ouvry, ``Interacting gauge fields of any spin and symmetry,''
  Phys. Lett. B {\bf 179} (1986) 115; Erratum-ibid. {\bf 183} B (1987) 434.
\bibitem{boulanger}
  Nicolas Boulanger, Serge Leclercq, Per Sundell,
  ``On The Uniqueness of Minimal Coupling in Higher-Spin Gauge Theory,''
  JHEP 0808:056,2008;  [arXiv:0805.2764 [hep-th]].
  Xavier Bekaert, Nicolas Boulanger, Sandrine Cnockaert, Serge Leclercq,
  ``On Killing tensors and cubic vertices in higher-spin gauge theories,''
  Fortsch. Phys. {\bf 54} (2006) 282-290; [arXiv:hep-th/0602092].
\bibitem{Petkou}
  A.~Fotopoulos, N.~Irges, A.~C.~Petkou and M.~Tsulaia,
  ``Higher-Spin Gauge Fields Interacting with Scalars: The Lagrangian Cubic
  Vertex,''
  JHEP {\bf 0710} (2007) 021;
  [arXiv:0708.1399 [hep-th]].
  I.~L.~Buchbinder, A.~Fotopoulos, A.~C.~Petkou and M.~Tsulaia,
  ``Constructing the cubic interaction vertex of higher spin gauge fields,''
  Phys.\ Rev.\  D {\bf 74} (2006) 105018;
  [arXiv:hep-th/0609082].
 Grav. 21 (2004) S1457;
\bibitem{review}
    M. A. Vasiliev, ``Higher Spin Gauge Theories in Various Dimensions'',
    Fortsch. Phys. 52, 702 (2004) [arXiv:hep-th/0401177].
    X. Bekaert, S. Cnockaert, C. Iazeolla and M. A. Vasiliev,
    ``Nonlinear higher spin theories in various dimensions'', [arXiv:hep-th/0503128].
    D. Sorokin,``Introduction to the Classical Theory of Higher Spins'' AIP Conf. Proc. 767, 172
    (2005); [arXiv:hep-th/0405069].
N. Bouatta, G. Compere and A. Sagnotti, ``An Introduction to Free Higher-Spin Fields''; [arXiv:hep-th/0409068].
\bibitem{Kleb}
  I.~R.~Klebanov and A.~M.~Polyakov, ``AdS dual of the critical O(N)
  vector model,'' Phys.\ Lett.\ B {\bf 550} (2002) 213;
  [arXiv:hep-th/0210114].
\bibitem{MMR1}
R.~Manvelyan, K.~Mkrtchyan and W.~R\"uhl,
  ``Ultraviolet behaviour of higher spin gauge field propagators and one loop
  mass renormalization,''
  Nucl.\ Phys.\  B {\bf 803} (2008) 405
  [arXiv:0804.1211 [hep-th]].
\bibitem{Frons}
    C.~Fronsdal, ``Singletons And Massless, Integral Spin Fields On De Sitter Space (Elementary
    Particles In A Curved Space Vii),'' Phys.\ Rev.\ D {\bf 20},
    (1979) 848;``Massless Fields With Integer Spin,'' Phys.\ Rev.\ D {\bf
18} (1978) 3624.
\bibitem{Ruehl}
  W.~R\"uhl, ``The masses of gauge fields in higher spin field theory on
  AdS(4),'' Phys.Lett. B {\bf 605} (2005) 413; [arXiv:hep-th/0409252]; the results
  presented here are based on extensive calculations performed by K. Lang and
  W.~R\"uhl, Nucl. Phys. B {\bf 400} (1993) 597.
\bibitem{MR1}
  R.~Manvelyan and W.~R\"uhl,
  ``The off-shell behaviour of propagators and the Goldstone field in  higher
  spin gauge theory on AdS(d+1) space,''
  Nucl.\ Phys.\  B {\bf 717} (2005) 3;
  [arXiv:hep-th/0502123].
 \bibitem{MR2}
 R.~Manvelyan and W.~R\"uhl,
  ``The masses of gauge fields in higher spin field theory on the bulk of
  AdS(4),''
  Phys.\ Lett.\  B {\bf 613} (2005) 197;
  [arXiv:hep-th/0412252].
 \bibitem{MR3}
  R.~Manvelyan and W.~R\"uhl,
  ``The structure of the trace anomaly of higher spin conformal currents in the
  bulk of AdS(4),''
  Nucl.\ Phys.\  B {\bf 751}, (2006) 285;
  [arXiv:hep-th/0602067].
\bibitem{MR4} R.~Manvelyan and W.~R\"uhl,
  ``The quantum one loop trace anomaly of the higher spin conformal  conserved
  currents in the bulk of AdS(4),''
  Nucl.\ Phys.\  B {\bf 733} (2006) 104;
  [arXiv:hep-th/0506185].
\bibitem{MR5}
  R.~Manvelyan and W.~R\"uhl,
  ``Generalized Curvature and Ricci Tensors for a Higher Spin Potential and the
  Trace Anomaly in External Higher Spin Fields in $AdS_{4}$ Space,''
  Nucl.\ Phys.\  B {\bf 796} (2008) 457;
  [arXiv:0710.0952 [hep-th]].
\bibitem{DF}
B.~deWit and D.Z.~Freedman, ``Systematics of higher spin gauge
fields,'' Phys. Review D \textbf{21} (1980), 358-367.
\bibitem{MR6}
  R.~Manvelyan and W.~R\"uhl,
  ``The Generalized Curvature and Christoffel Symbols for a Higher Spin
  Potential in $AdS_{d+1}$ Space,''
  Nucl.\ Phys.\  B {\bf 797}, 371 (2008)
  [arXiv:0705.3528 [hep-th]].
\end{thebibliography}
\end{document}